\documentstyle[12pt]{article}
\begin{document}

\begin{center}
{\small \bf DEUTERON ELECTRIC QUADRUPOLE AND OCTUPOLE\\  
POLARIZABILITIES} \\[.4in]

\setlength{\baselineskip}{0.1in}
{\footnotesize V. F. KHARCHENKO} \\[.05in]

{\footnotesize \it Bogolyubov Institute for Theoretical Physics,\\
National Academy of Sciences of Ukraine, UA - 03143, Kyiv, Ukraine\\
vkharchenko@bitp.kiev.ua} \\[.4in]
\end{center}

\begin{abstract}
\setlength{\baselineskip}{0.1in}
\noindent
{\footnotesize The direct transition-matrix approach to determination of the electric 
polarizabilities of quantum bound systems developed in my recent work$^1$ is
applied to study the electric multipole polarizabilities of a two-particle 
bound complex with a central interaction between the particles. Expressions for 
the electric quadrupole and octupole polarizabilities of the deuteron are derived 
and their values in the case of the $S$-wave separable interaction potential are 
calculated.} \\[.2in]
{\footnotesize \it Keywords}: {\footnotesize Electric multipole polarizabilities;
two-body quantum systems; deuteron} \\[.2in]
{\footnotesize PACS Nos.: 21.10.Ky; 21.45.+v; 27.10.+h;}
\end{abstract}

\vspace*{.1in}
\noindent {\bf 1. Introduction} \\ [.1in]
The electric polarizabilities of hadronic bound complexes are important 
physical quantities that characterize the deformation properties of composite 
structures containing information on the fundamental nuclear force between 
their constituents. 

The result of the direct determination of the electric dipole polarizability of 
the deuteron by measuring deviations from the Rutherford scattering for d-$^{208}\mbox{Pb}$ 
elastic scattering below the Coulomb barrier,
$$
\alpha_E(^2\mbox{H})=0.70 \pm 0.05\;\; \mbox{fm}^3\;(\mbox{Ref.2})\,,
$$
is near to the result deduced from data for the total $^2$H 
photoabsorption cross section using the $\sigma_{-2}$ sum rule, 
$$
\alpha_E(^2\mbox{H})=0.61 \pm 0.04\;\; \mbox{fm}^3\;(\mbox{Ref. 3})\,.
$$

The calculated values of the deuteron electric dipole polarizability, 
performed both with the simple separable potential that includes the 
tensor interaction between nucleons$^4$ ($\alpha_E(^2\mbox{H})=0.6311\; \mbox{fm}^3$) 
and with more realistic nucleon-nucleon potentials$^5$ 
($\alpha_E(^2\mbox{H})=0.6328\; \mbox{fm}^3$) are within the experimental error of 
the result of Ref. 3, both the data$^{2,3}$ being nearby.

The anisotropic character of the deuteron electric dipole polarizability 
caused by the presence of the tensor interaction between the neutron and proton 
has been explored both by the method of the traditional nuclear physics$^{4,6}$ 
and in the framework of the chiral effective field theory$^{7,8}$. As it was shown 
in Ref. 4, the longitudinal component (with the electric field along the deuteron 
spin) $\alpha^1_{E1}$ and the transverse component $\alpha^0_{E1}$ are essentially 
differed between themselves: $\alpha^1_{E1}=0.669\; \mbox{fm}^3$ and 
$\alpha^0_{E1}=0.555\; \mbox{fm}^3$. (The abovementioned electric dipole polarizability 
$\alpha_{E1}(^2\mbox{H}$) is an averaged value of the longitudinal and transverse 
components: $\alpha_{E1}=\frac{2}{3} \alpha^1_{E1}+\frac{1}{3} \alpha^0_{E1}$.) 

Unfortunately, up to now the deuteron electric polarizabilities of the higher 
multipole order have not been experimentally determined yet.

The influence of the deuteron polarization (distortion) in the processes of scattering 
and disintegration of the deuteron in the Coulomb field of a nucleus at low energies 
was earlier studied in the papers by Abelishvili and Sitenko$^9$ and by Clement$^{10}$. 

The aim of this work is to study theoretically the deuteron electric multipole 
polarizabilities by applying the direct transition-matrix approach formulated 
in the recent paper$^1$. In Section 2 the formula for the electric multipole 
polarizability of the two-body bound system composed of one charged and one neutral 
particles with a central interaction is derived. The use of the $t$-matrix approach 
in the case of the $S$-wave separable interaction potential is carried out in Section 3.
Section 4 is devoted to the results of calculation of the deuteron quadrupole and 
octupole polarizabilities and discussion of them. \\

\vspace*{.1in}
\noindent {\bf 2. Transition-matrix approach to determination of the multipole 
polarizabilities of a two-body bound complex} \\ [.1in]
The electric multipole polarizabilities of a bound complex $\alpha_{E\lambda}$ 
are the coefficients of the asymptotic expansion of the polarization potential 
of interaction between a charged particle $0$ and the center of mass of the complex 
at distances $\rho$ much more than the complex size,
\begin{equation}
V_{pol}(\rho_0) = - \frac{e_0^2}{2} \sum_{\lambda = 1}^{\infty} 
\frac{\alpha_{E\lambda}}{\rho_0^{2\lambda+2}}\;,
\end{equation}
where $e_0$ is the charge of the particle $0$. In the case of a two-body complex 
that consists of a charged particle $1$ and a neutral particle $2$ (for instance, 
the deuteron), the electric polarizability of the multipole order $2^{\lambda}$ 
is given by
\begin{equation}
\alpha_{E\lambda}=-2<\psi_0\mid M_{\lambda} g^Q(-b) M_{\lambda}\mid \psi_0>\,.
\end{equation}
Here $\psi_0$ is the wave function of the bound complex in the state with the 
energy $-b$, $g^Q(-b)$ is the "truncated" Green's operator at the energy $-b$,
\begin{equation}
g^Q(\varepsilon)=Q g(\varepsilon),\;\; g(\varepsilon)=(\varepsilon - h^{\circ}  
- v_{12})^{-1},\;\; Q=1-P,\;\; P=\psi_0><\psi_0\mid \;,
\end{equation}
$g(\varepsilon)$ is the total Green's operator, $P$ is the operator projecting 
on the ground state of the complex,  $h^{\circ}$ and $v_{12}$ are the kinetic 
energy and interaction potential operators of the relative motion of the constituent 
particles $1$ and $2$ of the complex, $M_{\lambda}$ is the multipole operator
\begin{equation}
M_{\lambda}=e_1 r_1^{\lambda} P_{\lambda}(\hat{{\bf r}}_1 \cdot \hat{{\mbox{\boldmath$\rho$}}})
= e_1 \left(\frac{m_2}{m_{12}}\right) r^{\lambda} P_{\lambda}(\hat{{\bf r}}\cdot 
\hat{{\mbox{\boldmath$\rho$}}})\;,
\end{equation}
where $e_1$ is the charge of the particle $1$ of the complex, ${\bf r}_i$ is the 
radius-vector of the particle $i$ of the complex relative to its center of mass, 
${\bf r}_1=\frac{m_2}{m_{12}}{\bf r}$,  ${\bf r}$ is the relative radius-vector 
between the particles of the complex, ${\bf r}={\bf r}_1-{\bf r}_2$, 
${\mbox{\boldmath$\rho$}}$ is the relative radius-vector between the particle $0$ 
and the center of mass of the complex, $m_i$ is the mass of the particle $i$, 
$m_{12}= m_1+m_2$, $P_{\lambda}(x)$ is the Legendre polynomial. The unit vectors 
in (4) are marked by the hat. 

In the momentum space (with  ${\bf r}=i\hbar\nabla_{\bf k}$) the function 
$<{\bf k}\mid M_{\lambda}\mid \psi_0>$ that is contained in Eq. (2) can be 
written as 
\begin{equation}
\langle {\bf k} \mid M_{\lambda} \mid \psi_0 \rangle = 
i^{\lambda} e_1 \left( - \frac{m_1}{m_{12}} \right)^{\lambda} \varphi_{\lambda}(k) 
P_{\lambda}(\hat{{\bf k}} \cdot \hat{{\mbox{\boldmath $\rho$}}}) \;,
\end{equation}
where
\begin{equation}
\varphi_{\lambda}(k)= (-1)^{\lambda} k^{\lambda}\left[ \left( \frac{1}{k} \frac{d}{dk} \right)^{\lambda} 
\psi_0(k) \right]\;. 
\end{equation}

By its definition (3), the "truncated" Green's operator $g^Q(\varepsilon)$ has no singularity 
at $\varepsilon=-b$. To show explicitly cancellations of singularities in Eq. (3), it 
is worthwhile to express the total Green's function in terms of the transition operator, 
\begin{equation}
g(\varepsilon)=g^{\circ}(\varepsilon)+g^{\circ}(\varepsilon)t(\varepsilon)g^{\circ}(\varepsilon). 
\end{equation}
Here $g^{\circ}(\varepsilon)=(\varepsilon - h^{\circ})^{-1}$ is the free Green's 
operator and $t(\varepsilon)$ is the transition operator satisfying the 
Lippmann-Schwinger equation
\begin{equation}
t(\varepsilon)=v+v g^{\circ}(\varepsilon)t(\varepsilon)\;. 
\end{equation}

In the case of a central interaction potential $v$, the operator $t(\varepsilon)$ 
can be decomposed into partial components $t_l(\varepsilon)$, each of them being  
characterized by a definite value of the orbital angular momentum of relative motion 
of particles $l$. The partial wave expansion of the transition matrix is given by
\begin{equation}
<{\bf k}|t^{(h)}(\varepsilon)|{\bf k^{\prime}}>=\sum_{l=1}^{\infty} (2l+1) 
t_l(k,k^{\prime};\varepsilon) P_{l}(\hat{\bf k} \cdot \hat{\bf k}^{\prime})\;.
\end{equation}
The $S$-wave partial component of the transition operator, $t_0(\varepsilon)$, has pole 
singularity at the point of the energy of the bound complex $\varepsilon=-b$ and can be 
represented as the sum of the pole $t_0^P$ and smooth  $\tilde{t}_0$ parts,
\begin{equation}
t_0(\varepsilon) = t_0^P(\varepsilon) + \tilde{t}_0(\varepsilon)\;,  
\end{equation}
\begin{equation}
t_0^P(\varepsilon) = \frac{\mid u_0><u_0\mid}{\varepsilon+b} 
\end{equation}
with the vertex function
\begin{equation}
\mid u_0\rangle = -[g^{\circ}(-b)]^{-1} \mid \psi_0 \rangle \;. 
\end{equation}

Cancelling the pole singularities of the operators $g(\varepsilon)$ and $Pg(\varepsilon)$ 
at $\varepsilon=-b$ in Eq. (3) we write the operator $g^Q(\varepsilon)$ as
\begin{equation}
\begin{array}{rcl}
g^Q(-b) = g^{\circ}(-b)& - &g^{\circ}(-b) \mid \psi_0 \rangle \langle \psi_0 \mid 
- \mid \psi_0 \rangle \langle \psi_0 \mid g^{\circ}(-b) \\[2mm]  \nonumber
& + &g^{\circ}(-b)\tilde{t}(-b) g^{\circ}(-b)\;.  
\end{array}
\end{equation}
Here the operator $\tilde{t}(-b)$ denotes the smooth part of the transition operator 
$t$, it is the sum of the $S$-wave component $\tilde{t}_0(-b)$ and all the higher 
(with $l\geq 1$) partial components of the transition operator $t(-b)$ that also 
have not singularities at this point,
\begin{displaymath}
\tilde{t}(-b) =\tilde{t}_0(-b) + t^{(h)}(-b)\;,  
\end{displaymath}
\begin{equation}
<{\bf k}|t^{(h)}(-b)|{\bf k^{\prime}}>=\sum_{l=1}^{\infty} (2l+1) t_l(k,k^{\prime}; -b) 
P_{l}(\hat{\bf k} \cdot \hat{\bf k}^{\prime})\;.
\end{equation}

One can readily see that the smooth $S$-wave operator $\tilde{t}_0(\epsilon)$ at 
the point $\varepsilon=-b$ is separable,
\begin{equation}
\tilde{t}_0(-b) = \mid u_0\rangle \left(-\frac{R_1}{b} \right) \langle u_0\mid\;,\;\; 
R_1 = b \langle u_0\mid (b+h^{\circ})^{-3} \mid u_0\rangle\;,
\end{equation}
with the formfactor in the form of the vertex function (12).

Substituting the expressions for the "truncated" Green's operator $g^Q(-b)$ (13) and for 
the smooth components of the transition operator $\tilde{t}(-b)$ and $\tilde{t}_0(-b)$, 
(14) and (15), into Eq. (2) and taking into account the relations
\begin{equation}
\begin{array}{rcl}
\langle\psi_0\mid M_{\lambda} \mid \psi_0\rangle\; & = & \; 
e_2 \delta_{\lambda 0}\;,
 \\[3mm] \nonumber
\langle\psi_0\mid g^{\circ}(-b)M_{\lambda}\mid \psi_0\rangle\; & = &\;
\langle\psi_0\mid M_{\lambda}g^{\circ}(-b)\mid \psi_0\rangle^{*}\; = \; 
-e_2 \frac{R_1}{b}\delta_{\lambda 0}\;,
\end{array}
\end{equation}
we write the electric multipole polarizability of the two-body complex in the form
\begin{equation}
\begin{array}{rcl}
\alpha_{E\lambda} & = & -\;\;2\langle\psi_0\mid M_{\lambda} g^{\circ}(-b)
M_{\lambda}\mid \psi_0\rangle \\[3mm] \nonumber
&   & -\;\;2\langle\psi_0\mid M_{\lambda} g^{\circ}(-b)t^{(h)}(-b) g^{\circ}(-b) 
M_{\lambda}\mid \psi_0\rangle\;,\quad \lambda=1,2,3,\cdots \;. 
\end{array}
\end{equation}

After integrating in Eq. (17) over angular variables with the use of the formula (5) 
for the function $M_{\lambda}\psi_0$ and the partial expansion (14) for the operator 
$t^{(h)}(-b)$, we find the final expression for the electric multipole polarizability 
of the two-body complex with a central interaction between the constituent particles: 
\begin{displaymath}
\alpha_{E\lambda} = \frac{2}{2\lambda+1}{e_1}^2
\left(\frac{m_2}{m_{12}}\right)^{2\lambda}
\left\{\int_{0}^{\infty} \frac{dk k^2}{2\pi^2} \frac{\mid \varphi_{\lambda}(k)
\mid^2}{\frac{k^2}{2\mu_{12}}+b}\right. \\ [1mm] 
\end{displaymath}
\begin{equation}
\left.- \int_{0}^{\infty}\int_{0}^{\infty} \frac{dk k^2 dk^{\prime}k^{{\prime}2}}
{4\pi^4}\frac{\varphi^{*}_{\lambda}(k) t_{\lambda}(k,k^{\prime};-b)
\varphi_{\lambda}(k^{\prime})}
{(\frac{k^2}{2\mu_{23}}+b)(\frac{k^{{\prime}2}}{2\mu_{23}}+b)}\right\}\;.  \\ [3mm] 
\end{equation}
Here, the functions $\varphi_{\lambda}$, which are built out of derivatives of 
the wave function of the ground state of the complex according to Eq. (6), 
for the dipole ($\lambda=1$), quadrupole ($\lambda=2$) and octupole 
($\lambda=3$) polarizabilities take the form
\begin{equation}
\varphi_{\lambda}(k) = \left\{ \begin{array}{lr}
 - \psi_0^{\prime}(k) & \mbox{for $\lambda=1$}\;, \\ 
 + \psi_0^{\prime\prime}(k) - \frac{1}{k} \psi_0^{\prime}(k) & \mbox{for $\lambda=2$}\;, \\
 - \psi_0^{\prime\prime\prime}(k) + \frac{3}{k} \psi_0^{\prime\prime}(k) 
 - \frac{3}{k^2} \psi_0^{\prime}(k) & \mbox{for $\lambda=3$}\;,
                                \end{array}
                      \right.                                
\end{equation}
and the corresponding partial components of the transition matrix satisfy the 
Lippmann-Schwinger equation
\begin{equation}
t_{\lambda}(k,k^{\prime};-b) = v_{\lambda}(k,k^{\prime})-
\int_{0}^{\infty} \frac{dk^{\prime\prime} k^{{\prime\prime}2}}{2\pi^2}
v_{\lambda}(k,k^{\prime\prime}) \frac{1}{\frac{k^{{\prime\prime}2}}{2\mu_{12}}+b}
t_{\lambda}(k^{\prime\prime},k^{\prime};-b)\;\;, \\[3mm]  
\end{equation}
where $v_{\lambda}(k,k^{\prime})$ is the partial component of the interaction 
potential $<{\bf k}|v|{\bf k^{\prime}}>$, $\mu_{12}$ is the reduced mass of the 
constituents.

The formula (18) is rigorous, it provides a possibility to calculate the multipole 
polarizabilities of both Coulomb and nuclear bound complexes with a central 
interaction between constituents. In accordance with Eqs. (18) and (19), to determine 
the $2^{\lambda}$-pole polarizability of a two-particle complex, it is necessary 
to know only lower-order derivatives of the momentum-space bound-state wave function  
and a single partial component of the transition matrix at the negative 
energy of the bound state. (The $P$-, $D$-, and $F$-wave partial components of the 
t-matrix are needed to determine the dipole, quadrupole, and octupole polarizabilities, 
respectively.)

Compared to the approach, which is traditionally applied to calculate the electric 
dipole polarizability of atomic and nuclear two-body systems using the spectral 
expansion of the Green's operator $g(-b)$ in the complete system of eigenfunctions 
of the Hamiltonian  $h=h^{\circ}+v_{12}$,
\begin{equation}
\alpha_{E1} = 2 \sum_{n \neq 0}^{} \frac{\mid \langle \psi_n \mid {\bf D}_1 \cdot 
\hat{\mbox{\boldmath$\rho$}}\mid \psi_0 \rangle \mid ^2}{{\epsilon}_n + b}\;,
\end{equation}
where the summation is over all the possible discrete and continuous excited states 
$n$ with the the energy ${\epsilon}_n$ and the wave function $\psi_n$, 
$t$-matrix approach (18) is much simpler and mathematically clear since it rests only on 
the bound-state wave function and the corresponding  partial component of the transition 
matrix at the negative energy of the state, both of them being real functions of relative 
momentum variables. With the use of the $t$-matrix approach, there is no need for 
calculation of more complicated complex continuum wave functions at all. \\

\vspace*{.1in}
\noindent {\bf 3. $S$-wave separable model} \\ [.1in] 
The expression (18) for the electric multipole polarizability of the two-body bound complex 
is obtained assuming that the interaction between constituents is central. 

A further simplification of the formula (18) for the polarizability of the complex can be 
obtained describing the interaction between the charged particle $1$ and the neutral 
particle $2$ of the complex with the use of the pure $S$-wave separable potential$^{11}$
\begin{equation}
v_0(k,k^{\prime})=-\nu u_0(k) u_0(k^{\prime})\;.
\end{equation}
In this case the wave function of the two-particle complex equals to
\begin{equation}
\psi_0(k)=\frac{u_0(k)}{\frac{k^2}{2\mu_{12}}+b}\;.
\end{equation} 
Using the Yukawa formfactor (Yu) in Eq. (22) 
\begin{equation}
u_0(k)=\frac{(2\pi)^{3/2}\cal N}{k^2+\beta^2}
\end{equation} 
and the relationship between the parameters $\nu$ and $\beta$ that takes place if the 
potential (22) forms one bound state of the complex with the binding energy 
$b=\frac{\kappa^2}{2\mu_{12}}$, 
\begin{equation}
\nu=\frac{\beta(\beta+\kappa)}{2\pi^2\mu_{12}} \;,
\end{equation} 
we write the normalized wave function of the bound complex as
\begin{equation}
\psi_0(k)=\frac{C_0}{(k^2+\kappa^2)(k^2+\beta^2)}\;,\quad 
C_0=2\mu_{12}(2\pi)^{3/2}{\cal N}=\sqrt{8\pi\kappa\beta(\beta+\kappa)^3}\;.
\end{equation} 
In this case, the functions $\varphi_{\lambda}(k)$ (19) formed from derivatives of 
the wave function (26) are given by  \\ [1mm]
\begin{displaymath}
\varphi_1(k)=2C_0\frac{k(2k^2+\beta^2+\kappa^2)}
{(k^2+\beta^2)^2(k^2+\kappa^2)^2}\;, 
\\ [1mm] 
\end{displaymath}
\begin{equation} 
\varphi_2(k)=8C_0\frac{k^2\{3k^4+3(\beta^2+\kappa^2)k^2+(\beta^4+
\beta^2\kappa^2+\kappa^4)\} }{(k^2+\beta^2)^3(k^2+\kappa^2)^3}\;, 
\\ [1mm]  
\end{equation}
\begin{displaymath}
\varphi_3(k)=48C_0\frac{k^3\{4k^6+6(\beta^2+\kappa^2)k^4+
4(\beta^4+\beta^2\kappa^2+\kappa^4)k^2+(\beta^2+\kappa^2)(\beta^4+\kappa^4)\} }
{(k^2+\beta^2)^4(k^2+\kappa^2)^4}\;. 
\\  [3mm]
\end{displaymath}

Note that the considered model (22) does not take account of the interactions in 
the higher orbital states. Also, the sole non-zero $S$-wave partial component of 
the transition matrix $t_0$ is not contained in the expression (18) for the 
multipole polarizabilities. Therefore, in this model the second term of the formula 
(18), in which $t_{\lambda}$ is present only with $\lambda > 0$, is equal to zero, 
and the expression (18) is reduced to \\ [1mm]
\begin{equation}
\alpha_{E\lambda} = \frac{4}{2\lambda+1}
\left(\frac{m_2}{m_{12}}\right)^{2\lambda+1}\frac{m_1 e_1^2}{\hbar^2}
\int_{0}^{\infty} \frac{dk k^2}{2\pi^2} \frac{\mid \varphi_{\lambda}(k)
\mid^2}{k^2+\kappa^2}\;. \\  [3mm]
\end{equation}

Substituting then the functions (27) into the formula (28) and performing integration 
we obtain the following expressions for the dipole, quadrupole and octupole 
polarizabilities of the two-particle complex:

\begin{equation}
\alpha_{E1}=\frac{1}{12} \left( \frac{m_2}{m_{12}} \right)^3 \frac{m_1e_1^2}{{\hbar}^2}
\frac{3\beta^5+18{\beta}^4\kappa+51{\beta}^3{\kappa}^2+96{\beta}^2{\kappa}^3+
48{\beta}{\kappa}^4+8{\kappa}^5}{{\beta}^2(\beta+\kappa)^3{\kappa}^4}\;,
\end{equation} 
\begin{displaymath}
\alpha_{E2}=\frac{1}{10} \left( \frac{m_2}{m_{12}} \right)^5 \frac{m_1e_1^2}{{\hbar}^2}
\frac{1}{{{\beta}^4(\beta+\kappa)^5{\kappa}^6}}
\{5\beta^9+40{\beta}^8\kappa+150{\beta}^7{\kappa}^2+360{\beta}^6{\kappa}^3 \qquad
\end{displaymath}
\begin{equation}
+645{\beta}^5{\kappa}^4+960{\beta}^4{\kappa}^5
+640{\beta}^3{\kappa}^6+332{\beta}^2{\kappa}^7+96{\beta}{\kappa}^8+12{\kappa}^9\}\;,
\end{equation} 
\begin{displaymath}
\alpha_{E3}=\frac{9}{112} \left( \frac{m_2}{m_{12}} \right)^7 \frac{m_1e_1^2}{{\hbar}^2}
\frac{1}{{{\beta}^6(\beta+\kappa)^7{\kappa}^8}}
\{35\beta^{13}+350{\beta}^{12}\kappa+1645{\beta}^{11}{\kappa}^2 \qquad \qquad
\end{displaymath}
\begin{equation}
+4900{\beta}^{10}{\kappa}^3+10605{\beta}^9{\kappa}^4+18270{\beta}^8{\kappa}^5
+26915{\beta}^7{\kappa}^6+35840{\beta}^6{\kappa}^7
\end{equation}
\begin{displaymath}
+26880{\beta}^5{\kappa}^8+18000{\beta}^4{\kappa}^9+9760{\beta}^3{\kappa}^{10}
+3616{\beta}^2{\kappa}^{11}+800{\beta}{\kappa}^{12}+80{\kappa}^{13}\}\;.      
\end{displaymath}    \\  

In the case of the zero-range interaction ($\beta\rightarrow\infty$) the expressions 
(29) - (31) simplify to the form
\pagebreak
\begin{displaymath}
\alpha_{E1}=\frac{1}{4} \left( \frac{m_2}{m_{12}} \right)^3 \frac{m_1e_1^2}{{\hbar}^2} 
\frac{1}{\kappa^4}\;,
\end{displaymath}
\begin{equation}
\alpha_{E2}=\frac{1}{2} \left( \frac{m_2}{m_{12}} \right)^5 \frac{m_1e_1^2}{{\hbar}^2}
\frac{1}{\kappa^6}\;,
\end{equation}
\begin{displaymath}
\alpha_{E3}=\frac{45}{16} \left( \frac{m_2}{m_{12}} \right)^7 \frac{m_1e_1^2}{{\hbar}^2}
\frac{1}{\kappa^8}\;.   
\end{displaymath}   \\  

Note that the expression for the electric dipole polarizability of the two-body bound 
complex with the separable interaction (29) is in accordance with the corresponding 
formulae obtained by employing the Dalgarno-Lewis perturbation technique$^{12}$ 
in the paper by Friar and Fallieros$^{13}$ and in the framework of the three-body 
formalism of the effective interaction between a charged particle and a complex 
in our papers$^{14-16}$. The t-matrix approach with $S$-wave separable interaction 
has been applied to determine the electric dipole polarizabilities of the triton 
and lambda hypertriton as two-cluster systems in the previous our works$^{17,18}$.  \\

\vspace*{.1in}
\noindent {\bf 4. Numerical results and discussion} \\ [.1in]
In this section we apply the formalism developed above to the deuteron 
as a bound complex composed of the proton (the particle $1$, $e_1=e_p,\; 
m_1=m_p,\; \hbar^2/m_p e_p^2=28.819893\;\mbox{fm}$ is the proton Bohr radius) 
and the neutron (the particle $2$, $m_2=m_n$) with the binding energy $b=\kappa^2/2\mu_{pn}$. 
We calculate the dipole, quadrupole and octupole polarizabilities of the deuteron 
both in the model of the $S$-wave separable potential (22) with the Yukawa formfactor (24) 
and in the zero-range limit ($\beta\rightarrow\infty$).

We find the parameters  of the separable potential (22) and (24), $\nu$ and $\beta$, by fitting them 
to the low-energy $p-n$ interaction data: the deuteron binding energy $b(^2\mbox{H})$,
\begin{equation}
b(^2\mbox{H})=2.224575(9)\;\mbox{MeV}\;\; \mbox{(Ref. 17)}
\end{equation}
and the triplet $p-n$ scattering length $^3a_{pn}$ or the asymptotic $S$-wave normalization 
of the deuteron wave function $A_{S}(^2\mbox{H})$,
\begin{equation}
^3a_{pn}=5.424(3)\;\mbox{fm}\; \mbox{(Ref. 18)}\;,\;\;
A_{S}(^2\mbox{H})=0.8845(8)\;\mbox{fm}^{-1/2}\; \mbox{(Ref. 19)}\;.
\end{equation}
In the case of the model (22) and (24) we have
\begin{displaymath}
^3a_{pn}=\frac{2(\beta+\kappa)^2}{\beta \kappa (2\beta+\kappa ) } \;,\; 
A_{S}(^2\mbox{H})=\frac{ \sqrt{2\beta \kappa (\beta+\kappa)}}{\beta-\kappa } \;.
\end{displaymath}

The results of the calculation are shown in Table 1. The existing experimental data for 
the deuteron dipole polarizability are mentioned in Section 1.

\pagebreak

{\footnotesize \tablename\hspace{2mm}1.\hspace{1mm} The electric dipole 
$\alpha_{E1}$, quadrupole $\alpha_{E2}$ and octupole $\alpha_{E3}$
polarizabilities of the deuteron, $\alpha_{E\lambda}(\mbox{H})$, 
calculated with the use of the separable interaction potential (22) (Yu)  
and the zero-range interaction model (ZR) with the parameters fitting 
to the $p-n$ data (33) - (34)
\begin{center} \begin{tabular}{|c|c|c|c|} \hline
\multicolumn{1}{|c|}{}& 
\multicolumn{1}{c|}{}& 
\multicolumn{1}{c|}{}&
\multicolumn{1}{c|}{} \\
\multicolumn{1}{|c|}{$p-n$ interaction}&
\multicolumn{1}{c|}{ZR $\left[b(^2\mbox{H})\right]$}&
\multicolumn{1}{c|}{Yu $\left[b(^2\mbox{H}),\;^3a_{pn}\right]$}&
\multicolumn{1}{c|}{Yu $\left[b(^2\mbox{H}),\; A_{S}(^2\mbox{H})\right]$} \\
\multicolumn{1}{|c|}{$\left[ \mbox{parameters} \right] $}&
\multicolumn{1}{c|}{}&
\multicolumn{1}{c|}{}&
\multicolumn{1}{c|}{}\\ 
\multicolumn{1}{|c|}{}&
\multicolumn{1}{c|}{}&
\multicolumn{1}{c|}{}&
\multicolumn{1}{c|}{}\\ \hline
\multicolumn{1}{|c|}{}&
\multicolumn{1}{c|}{}&
\multicolumn{1}{c|}{}&
\multicolumn{1}{c|}{} \\ 
$\alpha_{E1}(^2\mbox{H}),\mbox{fm}^3$&$0.3776$&$0.6259$&$0.6292$ \\
\multicolumn{1}{|c|}{}&
\multicolumn{1}{c|}{}&
\multicolumn{1}{c|}{}&
\multicolumn{1}{c|}{} \\ 
$\alpha_{E2}(^2\mbox{H}),\mbox{fm}^5$&$3.5247$&$5.9129$&$5.9462$ \\
\multicolumn{1}{|c|}{}&
\multicolumn{1}{c|}{}&
\multicolumn{1}{c|}{}&
\multicolumn{1}{c|}{} \\ 
$\alpha_{E3}(^2\mbox{H}),\mbox{fm}^7$&$92.529$&$155.38$&$156.26$ \\ 
\multicolumn{1}{|c|}{}&
\multicolumn{1}{c|}{}&
\multicolumn{1}{c|}{}&
\multicolumn{1}{c|}{} \\  \hline
\end{tabular}
\end{center}}
\bigskip

It is known$^{13}$ that for the determination of the electric dipole polarizability 
of the deuteron the adequate description of the asymptotic behaviour of the deuteron 
wave function is of prime importance, since the weakly bound nucleons in the nucleus 
are with a great probability located at distances large compared to the interaction range. 
For the parameters of the potential fitted to both the $p-n$ bound-state data
($b(^2\mbox{H})$ and $A_{S}(^2\mbox{H})$), we find in Table 1 the following values of the 
deuteron electric polarizabilities
\begin{equation}
\alpha_{E1}(^2\mbox{H})=0.6292\;\mbox{fm}^3\;,\;\; 
\alpha_{E2}(^2\mbox{H})=5.9462\;\mbox{fm}^5\;,\mbox{ and } 
\alpha_{E3}(^2\mbox{H})=156.26\;\mbox{fm}^7\;. 
\end{equation}
The polarizabilities (35) are a little larger (by $0.5\;$\%
)than the ones obtained with the parameters fitted to the data $b$ and $^3a_{pn}$ (when 
the potential (22) reproduces a somewhat lesser value of the asymptotic normalization: 
$A_S=0.8820\;\mbox{fm}^{-1/2}$).

The deuteron dipole polarizability $\alpha_{E1}(^2\mbox{H})$ is further changed if the 
tensor $p-n$ interaction is taken into account. Thus, for example, with the use of 
separable tensor interaction potential$^{22,13}$ (that reproduces $A_S=0.8843\;\mbox{fm}^{-1/2}$) 
the dipole polarizability $\alpha_{E1}(^2\mbox{H})$ increases to $0.6311\;\mbox{fm}^3$ 
(Ref. 4) approaching the values obtained with the realistic potentials
$\alpha_{E1}(^2\mbox{H})=0.6328(17)\;\mbox{fm}^3$ (Ref. 5). As for the inclusion of 
the interaction in the higher partial states, notice that for the dipole polarizability 
of the deuteron $\alpha_{E1}(^2\mbox{H})$ it is more important the consideration of the 
tensor $p-n$ interaction (in $^3S_1+^3D_1$ eigenstate) than that of the interaction in the 
$P$-wave state$^{13}$. The tensor $p-n$ interaction also causes the anisotropy effect of 
the deuteron electric polarization---the dependence of the polarizability 
$\alpha^M_{E\lambda}(^2\mbox{H})$ on the projection of the total angular momentum 
(the spin) of the deuteron $M$ that describes the direction of the spin relative 
to the electric field vector. The anisotropy effect of the electric dipole polarizability 
has been previously studied in Ref. 4. The application of the $t$-matrix approach 
to the study of the anisotropy of the induced electric quadrupole moment of the 
deuteron having also its own quadrupole moment ($Q=0.2859(3)\;\mbox{fm}^2$ (Refs. 23 and 24)) 
still remains to be carried out. 

The advantage of the $t$-matrix formalism are clearly evident by the example of its 
application to the determination of the electric dipole polarizability of a simple 
two-particle system with long-range Coulomb interaction---the hydrogen atom. 
In this case, provided that the ratio of the electron mass to the proton mass tends 
to zero, the exact analytical value of the polarizability is known$^{25-27}$: 
$\alpha_{E1}(\mbox{H})=\frac{9}{2}a_B^3\;,\;\;a_B=\hbar^2/m_e e^2$ is the electron Bohr radius. 
With this, the formula (18) may be directly applied to the hydrogen atom. Substituting 
into the expression (18) the explicit analytical expression for the partial (with $l=1$) 
component of the Coulomb transition matrix at negative energy $-b$ obtained previously 
in Ref. 28 we find calculating separately the first (analytically) and second (numerically) 
terms that
$$
\alpha_{E1}(\mbox{H})=\frac{7}{3}a_B^3 + 0.321066940\;\AA^3=0.666831341\;\AA^3=4.5\;a_B^3\;.
$$
As distinct from the polarizability of the two-particle nuclear system, for which in Eq. (18) 
the first term containing the free propagator is dominant as compared to the second one, 
the polarizability of the Coulomb system (with the long-range interaction) is determined by 
the sum of the two very nearly equal terms---with the free propagator and with the 
$P$-wave partial component of the transition matrix at the negative energy.

The developed approach to the problem of the determination of the electric multipole 
polarizabilities, which is based on the two-particle transition matrix, is found to be rather 
effective as compared with the traditional method that uses the spectral expansion of the 
Green's operator of the system. The results by Castillejo et al$^{29}$ testify that with the 
use of the spectral expansion all the virtual excited bound states contribute $81.4\%$ 
of the magnitude of the electric multipole polarizability of the hydrogen atom, the rest 
is accounted for by the states of the continuum. With the application of th $t$-matrix 
approach the necessity of considering a large number of virtual states no longer arises 
at all. In the case of the dipole polarizability ($\lambda=1$), beside the term with the free 
virtual propagator, the formula (18) contains only one term with the $P$-wave partial 
transition $t$-matrix at the negative energy. Moreover, the transition matrix at negative 
energies is a real quantity, as opposed to more complicate complex functions of the continuum, 
which are used in the traditional approach.

Existing advantages of the $t$-matrix approach motivate necessity of its further 
development with the aim of applying it for two-particle systems with noncentral 
interactions (the tensor $n-p$ interaction) to study the anisotropic properties 
of the dynamic quadrupole polarization of the deuteron. Also, the extension of 
the transition-matrix formalism to investigate the polarizabilities of more intricate 
three- and four-body nuclear systems based on the Faddeev and Faddeev-Yakubovsky 
integral equations is under way. \\

\vspace*{.1in}
\noindent {\bf Acknowledgment} \\[.1in]
Partial support from the program "Development and application of the quantum field theory 
methods in nuclear, particle and condensed-matter physics" 
of the Physics and Astronomy Branch of the National Academy of Sciences of
Ukraine is gratefully acknowledged here. \\

\vspace*{.1in}
\noindent {\footnotesize {\bf References}
\vspace*{.1in}
\begin{itemize}
\setlength{\baselineskip}{.1in}
\item[{\tt 1.}]V. F. Kharchenko, nucl-th/1208.1394 .
\item[{\tt 2.}]N. L. Rodning, L. D.Knutson, W. G. Lynch, and M.
           B. Tsang, {\it Phys. Rev. Lett.} {\bf 49}, 909 (1982).
\item[{\tt 3.}]J. L. Friar, S. Fallieros, E. L. Tomusiak, D. Skopik and
           E. G. Fuller, {\it Phys. Rev.} {\bf C27}, 1364 (1983).               
\item[{\tt 4.}]A. V. Kharchenko, {\it Nucl. Phys.} {\bf A617}, 34 (1997).           
\item[{\tt 5.}]J. L. Friar and G. L. Payne, {\it Phys. Rev.} {\bf C55}, 2764
           (1997).
\item[{\tt 6.}]M. H. Lopes, J. A. Tostevin and R. C. Johnson, {\it Phys. Rev.} 
           {\bf C28}, 1779 (1983). 
\item[{\tt 7.}]J.-W. Chen, H. W. Grie{\ss}hammer, M. J. Savage and
           R. P. Springer, {\it Nucl. Phys.} {\bf A644}, 221 (1998);
           nucl-th/9806080. 
\item[{\tt 8.}]X. Ji and Y. Li, {\it Phys. Lett.} {\bf B591}, 76 (2004). 
\item[{\tt 9.}]T. L. Abelishvili and A. G. Sitenko, {\it Ukrainian J. Phys.}
           {\bf 6}, 3 (1961). 
\item[{\tt 10.}]C. F. Clement, {\it Phys. Rev.} {\bf 128}, 2724, 2728 (1962).            
\item[{\tt 11.}]Y. Yamaguchi, {\it Phys. Rev.} {\bf 95}, 1628 (1954). 
\item[{\tt 12.}]A. Dalgarno and J. T. Lewis, {\it Proc. Roy. Soc. (London)} {\bf A233}, 
           70 (1955). 
\item[{\tt 13.}]J. L. Friar and S. Fallieros, {\it Phys. Rev.} {\bf C29},
           232 (1984).           
\item[{\tt 14.}]V. F. Kharchenko, S. A. Shadchin and S. A. Permyakov, {\it
           Phys. Lett.} {\bf B199}, 1 (1987).
\item[{\tt 15.}]V. F. Kharchenko and S. A. Shadchin, {\it Three-body
           theory of the effective interaction between a particle and a
           two-particle bound system}, preprint ITP-93-24E (Institute for
           Theoretical Physics, Kyiv, 1993). 
\item[{\tt 16.}]V. F. Kharchenko and S. A. Shadchin, {\it Ukrainian J. Phys.}
           {\bf 42}, 912 (1997). 
\item[{\tt 17.}]V. F. Kharchenko and A. V. Kharchenko, {\it Collected Physical
           Papers (Lviv)} {\bf 7}, 432 (2008); nucl-th/0811.2565. 
\item[{\tt 18.}]V. F. Kharchenko and A. V. Kharchenko, {\it Int. J. Mod. Phys.} {\bf E19}, 
           225 (2010); nucl-th/1003.5769. 
\item[{\tt 19.}]C. Van der Leun and C. Alderliesten, {\it Nucl. Phys.}
           {\bf A380}, 261 (1982).
\item[{\tt 20.}]L. Koester and W. Nistler, {\it Z. Phys.} {\bf A272}, 189
           (1975). 
\item[{\tt 21.}]J. J. de Swart, C. P. F. Terheggen and V. G. J. Stoks,
           {\it Proc. of the Third Int. Symposium "Dubna Deuteron 95"},
           Dubna, Russia, 1995; nucl-th/9509032; J. J. de Swart, R. A. M.
           Klomp, M. C. M. Rentmeester and Th. A. Rijken,
           {\it Few-Body Syst. Suppl.} {\bf 99} (1995).                                         
\item[{\tt 22.}]Y. Yamaguchi and Y. Yamaguchi, {\it Phys. Rev.} {\bf 95}, 1635 (1954). 
\item[{\tt 23.}]R. V. Reid Jr. and M. L. Vaida, {\it Phys. Rev. Lett.} {\bf 29}, 494 (1972); 
           {\bf 34}, 1064 (1975)(E); {\it Phys. Rev.} {\bf A7}, 1841 (1973).
\item[{\tt 24.}]D. M. Bishop and L. M. Cheung, {\it Phys. Rev.} {\bf A20}, 381 (1979).
\item[{\tt 25.}]G. Wentzel, {\it Z. Phys.} {\bf 38}, 518 (1926).
\item[{\tt 26.}]I. Waller, {\it Z. Phys.} {\bf 38}, 635 (1926).
\item[{\tt 27.}]P. S. Epstein, {\it Phys. Rev.} {\bf 28}, 695 (1926).
\item[{\tt 28.}]S. A. Shadchin and V. F. Kharchenko, {\it J. Phys.} 
           {\bf 16}, 1319 (1983). 
\item[{\tt 29.}]L. Castillejo, I. C. Percival and M. J. Seaton, {\it Proc. Roy. Soc. 
           (London)} {\bf A254}, 259 (1960).  

\end{itemize}}

\end{document}